\documentclass[english,aps,pra,twocolumn,showpacs,superscriptaddress,floatfix]{revtex4}
\usepackage[T1]{fontenc}
\usepackage[latin1]{inputenc}
\usepackage{amsmath}
\usepackage{graphicx}
\usepackage{array}
\usepackage{amssymb}

\makeatletter

\usepackage{amsfonts}
\usepackage{epsfig}
\usepackage{dsfont}

\newcommand{\Eq}[1]{Eq.~(\ref{#1})}

\newcommand{\be}{\begin{equation}}
\newcommand{\bea}{\begin{eqnarray}}
\newcommand{\eea}{\end{eqnarray}}
\newcommand{\ee}{\end{equation}}

\def\H{{\cal H}}

\def\B{{\cal B}}
\def\A{{\cal A}}
\def\F{{\cal F}}

\newcommand{\UUU}{{\cal U}}

\def\one{\ensuremath{\hbox{$\mathrm I$\kern-.6em$\mathrm 1$}}}

\usepackage{babel}
\makeatother
\begin{document}

\title{Constrained optimization of sequentially generated entangled multiqubit states}

\author{Hamed Saberi}
\affiliation{Physics Department, Arnold Sommerfeld Center for Theoretical Physics, and Center for NanoScience,
Ludwig-Maximilians-Universit\"{a}t, Theresienstr.~37, 80333 Munich, Germany}

\author{Andreas Weichselbaum}
\affiliation{Physics Department, Arnold Sommerfeld Center for Theoretical Physics, and Center for NanoScience,
Ludwig-Maximilians-Universit\"{a}t, Theresienstr.~37, 80333 Munich, Germany}

\author{Lucas Lamata}
\affiliation{Max-Planck-Institut f\"ur Quantenoptik, Hans-Kopfermann-Strasse 1, 85748 Garching, Germany}

\author{David P\'erez-Garc\'{\i}a}
\affiliation{Dpto. An\'alisis Matem\'atico, Universidad Complutense de Madrid, 28040 Madrid, Spain}

\author{Jan \surname {von Delft}}
\affiliation{Physics Department, Arnold Sommerfeld Center for Theoretical Physics, and Center for NanoScience,
Ludwig-Maximilians-Universit\"{a}t, Theresienstr.~37, 80333 Munich, Germany}

\author{Enrique \surname {Solano}}
\affiliation{Departamento de Qu\'imica F\'isica, Universidad del Pa\'is Vasco--Euskal Herriko Unibertsitatea, Apdo. 644,
48080 Bilbao, Spain  \\ and IKERBASQUE, Basque Foundation for Science, Alameda Urquijo 36, 48011 Bilbao, Spain}

\date{September 20, 2009}

\begin{abstract}
We demonstrate how the matrix-product state formalism provides a flexible
structure to solve the {\it constrained} optimization problem associated with the sequential generation
of entangled multiqubit states under experimental restrictions. We consider a realistic scenario in which an ancillary system with
a limited number of levels performs restricted sequential interactions with qubits in a row. The proposed method relies on a suitable local optimization procedure, yielding an efficient recipe for the realistic and approximate sequential generation of any entangled multiqubit state. We give paradigmatic examples that may be of interest for theoretical and experimental developments.
\end{abstract}

\pacs{03.67.Bg, 02.70.--c, 42.50.Dv, 71.27.+a}

% Quantum state engineering and measurements,
% Computational techniques; simulations,
% Strongly correlated electron systems; heavy fermions

\maketitle

\section{Introduction}
\label{sec:intro}

Entangled multiqubit states are of central importance in the fields of
quantum computation and quantum communication~\cite{Nielsen_Chuang}, and have been the subject
of intensive theoretical and experimental investigations. As pointed out by Sch\"on \emph{et al.}~\cite{Schoen_PRL,Schoen_PRA}, the
classes of all sequentially generated multiqubit states, assisted by an itinerant ancilla, are exactly given by
the hierarchy of matrix-product states (MPSs)~\cite{David}. In this context, the required number of ancilla levels is determined by the dimension of the MPS canonical representation of the target multiqubit state. Matrix-product states play an important role in the context of strongly correlated systems~\cite{AKLT+Fannes} and describe the approximate ground states produced by density-matrix renormalization group (DMRG)
\cite{DMRG,DMRG_MPS} and Wilson's numerical renormalization group
\cite{Krishna,SWV_cloning}. Paradigmatic multiqubit states, such as Greenberger-Horne-Zeilinger (GHZ)~\cite{GHZ},
$W$~\cite{Warticle}, and cluster~\cite{Cluster} states, can be described by low-dimensional MPS and are considered valuable resources for quantum information and communication tasks.

The generation of multiqubit entangled states via a single global unitary operation acting on initially decoupled qubits is in general a difficult problem. From this point of view, several theoretical and experimental efforts have been oriented towards the sequential generation of paradigmatic entangled multipartite states. As a matter of fact, a number of sequential and global approaches have been implemented in different physical systems to produce specifically GHZ ~\cite{Pan00,Lei05}, $W$~\cite{Rau00,Haf05,Kie07}, and cluster~\cite{Kie05} states. In order to generate sequentially any multiqubit state, a wide range of ancilla levels and ancilla-qubit operations are necessary~\cite{Schoen_PRL}. In this sense, two important theoretical and experimental questions appear naturally: will the sequential generation of a desired multiqubit state still be feasible under given restricted experimental conditions? And if the answer is no, can we design an efficient protocol that tells us the best possible
approximation to the sequential generation of such a state? In this paper, we answer satisfactorily both questions. We demonstrate how the MPS formalism allows us to exploit linear algebraic tools to study this relevant constrained optimization problem~\cite{ConstrainedOptimizationBook}.

\section{Restrictions on the number of ancilla levels}
\label{sec:restricted_ancilla_levels}

%\emph{Restrictions on the number of ancilla levels.}
It is known that any $n$-qubit state $|\psi\rangle$ can be written canonically as an MPS with minimal dimension $D(\le 2^{n})$~\cite{David}.
It was also shown that such a state can be built sequentially with a $D$-dimensional ancilla, if we have access to arbitrary ancilla-qubit unitaries~\cite{Schoen_PRL}.
In the sequential generation of states, an ancillary system $\A$ (e.g., a $D$-level atom) couples sequentially to an initially decoupled qubit chain $|\psi_I\rangle=|\psi_I^{[n]}\rangle \otimes \dots \otimes |\psi_I^{[1]}\rangle$
(e.g., cavity photonic qubits that leak out after interacting with an atom). Assuming that in the last step
the ancilla decouples unitarily from the multiqubit system, we are left with the $n$-qubit state~\cite{Schoen_PRL}
\vspace{-3mm}
\begin{eqnarray}
\label{eq:MPS_nqubit}
|\psi\rangle=\sum_{i_n, \dots, i_1=0}^1 \langle\varphi_F| V_{[n]}^{i_n} \dots
V_{[1]}^{i_1} |\varphi_I\rangle |i_n,\dots,i_1\rangle   \; ,
\end{eqnarray}
an MPS of bond dimension $\dim({|\psi\rangle})=D$, where the $(D\times D)$-dimensional matrix $V_{[k]}^{i_k}$ represents the ancilla-qubit operation at step $k$ of the sequential generation (with isometry condition $\sum_{i_{k}=0}^1 V^{{i_k}\dagger} V^{i_k}=\mathds{1}$), with $|\varphi_I\rangle$ and $|\varphi_F\rangle$ being
the initial and the final ancilla states, respectively.
Hence, a relevant experimental question may be raised:
how well can we represent a given multiqubit state $|\psi\rangle$ if only an ancilla with a smaller number of levels, $D'<D$, is available? More formally: given a state $|\psi\rangle$, with a canonical MPS representation of bond dimension $D$, what is the optimal MPS $|\tilde{\psi}\rangle$ of
lower bond dimension $D'<D$ that minimizes their distance? We want to estimate
\begin{eqnarray}
\label{eq:MPS_smaller_D}
\min_{\dim({|\tilde{\psi}\rangle})=D'<D  }
\parallel |\psi\rangle - |\tilde{\psi}\rangle \parallel^{2}  \; .
\end{eqnarray}
We propose two techniques to perform the MPS approximation above, both exploiting a suitably designed local
optimization of the $V$ matrices in \Eq{eq:MPS_nqubit}. In the first approach, we make use of
a corollary of the singular value decomposition (SVD) theorem from linear algebra to perform a local
optimization procedure which may be called ``MPS compression,'' in analogy to the image compression technique
already used in computer science and engineering~\cite{SVD_image}. Let the SVD of matrix $A$ with $\mathrm{rank}(A)=r$ be
given by $A=\sum_{i=1}^{r} \sigma_i u_i v_i^{\dagger}$. Then, the best possible lower-rank approximation to $A$ that minimizes the Frobenius-norm distance $\min_{\mathrm{rank}(\tilde{A})=r'<r} {\|A-\tilde{A}\|}_{F}$
is given by $\tilde{A}=\sum_{i=1}^{r'} \sigma_i u_i v_i^{\dagger}$~\cite{Golub_VanLoan,Horn_Johnson}.
This suggests a \emph{truncation} scheme in which one keeps only the $r'$ largest singular values of $A$
to form the optimal lower-rank matrix $\tilde{A}$.
We exploit now this property, valid for a {\it single} matrix, and apply the outlined truncation to each matrix $V_{[k]}^{i_k}$, $k=1, ..., n$, in \Eq{eq:MPS_nqubit}, yielding an MPS of lower bond dimension $D'=D-(r-r')$.
This method offers a good solution for matrices with well-decaying singular-value spectrum.

In the second approach~\cite{SWV_cloning}, a DMRG-inspired variational optimization of $V$ matrices~\cite{VPC},
we seek the best possible approximation
to $|\psi\rangle$ in the space of all MPS $|\tilde{\psi}\rangle$ of the form~(\ref{eq:MPS_nqubit}) (with $V \to \tilde V$) with bond dimension
$D'<D$, by solving the minimization problem of \Eq{eq:MPS_smaller_D}
under the constant-norm condition $\langle \tilde{\psi}|\tilde{\psi}\rangle=1$, which is implemented using a Lagrange multiplier $\lambda$. Varying \Eq{eq:MPS_smaller_D}
with respect to the matrices defining $|\tilde{\psi}\rangle$ leads to a set of equations,
one for each $i_k$, of the form
\begin{eqnarray}
\label{eq:cloning_var}
\frac{\partial}{\partial \tilde{V}^{i_k}_{[k]}} \Bigl[(1+\lambda) \:
\langle \tilde{\psi} |  \tilde{\psi} \rangle
- 2 {\rm Re} \bigl( \langle \psi | \tilde{\psi}  \rangle \bigr)  \Bigr]=0  \; ,
\end{eqnarray}
which determines the optimal $\tilde{V}$ matrices of the desired state $|\tilde{\psi}\rangle$.
These equations can be solved very efficiently using a ``sweeping procedure'' in which
one fixes all but the $k$th $\tilde{V}$-matrix and solves the
corresponding Eq.~(\ref{eq:cloning_var}) for the matrix $\tilde{V}_{[k]}^{i_k}$.
Then one moves on to the neighboring site and, in this fashion, sweeps back and forth
through the chain until the convergence is reached.

\begin{figure}[t]
\centering
\includegraphics[width=.9\linewidth]{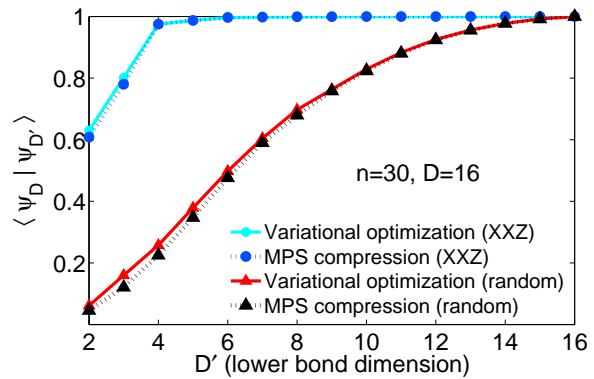}
\caption{(Color online) Comparison of the variational optimization approach (solid lines) with the MPS compression technique (dotted lines). We consider the ground state of the $XXZ$ Heisenberg Hamiltonian (circles) and a randomly initialized MPS (triangles), indicating how well these MPS with bond dimension $D$ can be approximated with those of dimension $D'<D$.
}
\label{MPS_approx}
\end{figure}

Figure~\ref{MPS_approx} illustrates the two  optimization schemes outlined above
for two different states, both with $D=16$, namely (i) the ground state of the $XXZ$ Heisenberg Hamiltonian and (ii) a randomly chosen MPS. For (i), which has a well-decaying singular-value spectrum, the ancilla dimension can be effectively reduced from 16 to 6. Since variational optimization allows for the feedback of information by several sweeps, it generally performs better than MPS compression.

\section{Restrictions on the source-qubit interactions}
\label{sec:restricted_unitaries}
%\emph{Restrictions on the source-qubit interactions.}
Every open-boundary MPS of form~(\ref{eq:MPS_nqubit}) (with $V \to A$)
%\begin{eqnarray}
%\label{eq:given_MPS}
%|\psi\rangle=\sum_{i_n \dots i_1=0}^1 \langle\varphi_F|
%A_{[n]}^{i_n} \dots A_{[1]}^{i_1} |\varphi_I\rangle |i_n,\dots,i_1\rangle  \;,
%\end{eqnarray}
with {\it arbitrary} $A$ matrices, not necessarily isometries, can be cast into a canonical MPS representation with minimal dimension $D$~\cite{Vidal}. Such states, as mentioned above \Eq{eq:MPS_nqubit}, can be generated sequentially ~\cite{Schoen_PRL}, such that the ancilla decouples unitarily in the last step. We note that the sequential generating isometries can be constructed explicitly
by successive SVD of the $A$ matrices and exploiting
the gauge freedom of the matrix-product states as outlined in Refs.~\cite{Schoen_PRL,Schoen_PRA}.
This is a general recipe for the sequential
generation of an~\emph{arbitrary} entangled multiqubit state if the required ancilla dimension $D$ and ancilla-qubit unitaries are available. However, in general, a given physical setup may not have access to some of the required local ancilla-qubit unitaries.
Given such a limitation, we face an interesting constrained optimization problem: which is the sequential protocol by which
a given multiqubit ``target'' state can be approximately generated with a maximal fidelity?

To address this problem, let us begin by considering the general unrestricted case:
the unitary time evolution of the joint system ancilla-qubit at step $k$ of the sequential generation may be described by a general unitary $U_{[k]}^{\A\B}:\H_\A\otimes\H_\B
\to \H_\A\otimes\H_\B$, $U_{[k]}^{\A\B}=e^{-iH_{[k]}^{\A\B} t/\hbar}$,
where $H_{[k]}^{\A\B}$ is a general bipartite Hamiltonian that \emph{couples} the
ancilla with the $k$th qubit. The latter can be written as
%\vspace{-4mm}
%\begin{eqnarray}
%\label{eq:entangling_Hamiltonian}
$H_{[k]}^{\A\B}=\sum_{j_{\A},j_{\B}=0}^{3} h_{j_{\A} j_{\B}}^{[k]} \sigma_{j_{\A}} \otimes \sigma_{j_{\B}}$ % \; ,
%\end{eqnarray}
where $h_{j_{\A} j_{\B}}^{[k]}$ are real-valued coupling constants and $\sigma_1$, $\sigma_2$,
$\sigma_3$ are the usual Pauli $\sigma$ matrices, with $\sigma_0\equiv I$ as the identity matrix. For the sake of simplicity, we have considered the case $D=2$, but similar generators can be found for $D>2$.

\begin{figure}[t]
\centering
\includegraphics[width=.95\linewidth]{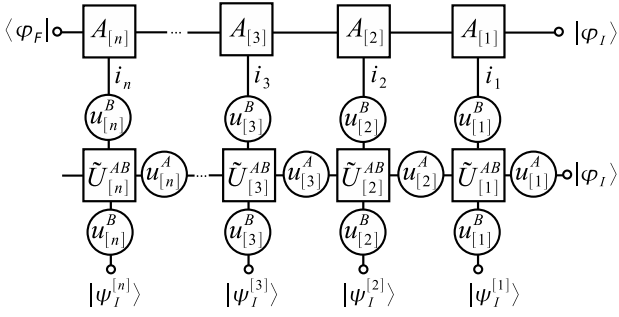}
\caption{The contraction pattern used to calculate the cost function in~\Eq{eq:cost_func} including the local
ancilla operations $\UUU^{\A}$ and local qubit operations $\UUU^{\B}$. The initial states of the qubits are denoted
by $|\psi_I^{[k]}\rangle$.}
\label{opt_pat}
\end{figure}

Now, suppose that only a restricted set of unitaries are available. As an illustrative case, let the entangling Hamiltonian have the restricted form of the $XY$ model~\cite{Whaley03}
\vspace{-3mm}
\begin{eqnarray}
\label{eq:restricted_Hamiltonian}
\tilde{H}_{[k]}^{\A\B}=h_1^{[k]} (\sigma_1 \otimes \sigma_1+\sigma_2 \otimes \sigma_2) \; ,
\end{eqnarray}
containing a single nonzero contribution $h_1^{[k]} \equiv h_{11}^{[k]}=h_{22}^{[k]}$.
Given an arbitrary MPS of the form of Eq.~(\ref{eq:MPS_nqubit}) (with $V \to A$) with arbitrary $A$ matrices and the restricted Hamiltonian of \Eq{eq:restricted_Hamiltonian}, the aim is to find the optimal restricted unitary operations ${\tilde{U}}_{[k]}^{\A\B}=e^{-i {\tilde{H}_{[k]}^{\A\B}} t/\hbar}$ that when applied sequentially to
an arbitrary initial state of the joint system $|\Phi_I\rangle=|\varphi_I\rangle \otimes |\psi_I\rangle$, yield a state of the form
\begin{eqnarray}
\label{eq:joint_psi}
|\tilde{\Psi}\rangle=\tilde{U}_{[n]}^{\A\B}\dots \tilde{U}_{[2]}^{\A\B} \tilde{U}_{[1]}^{\A\B} |\Phi_I\rangle \; ,
\end{eqnarray}
which is ``closest'' to the target state of the form $|\varphi_F \rangle \otimes | \psi \rangle$, where $|\varphi_F \rangle$ is arbitrary.
Note that the action of each restricted unitary on initial state of qubit, $\tilde{U}_{[k]}^{\A\B} |\psi_I^{[k]}\rangle$,
produces a restricted isometry of the form
\begin{eqnarray}
\label{eq:unitary_to_isometry}
\sum_{i_k,j_k,\alpha,\beta} \tilde{U}_{\alpha,\beta}^{i_k,j_k} |\alpha i_k\rangle \langle \beta j_k| \psi_I^{[k]}\rangle
=\sum_{i_k,\alpha,\beta} \tilde{V}_{\alpha,\beta}^{i_k}|\alpha i_k\rangle \langle \beta| \;,
\end{eqnarray}
with the definition $\tilde{V}_{\alpha,\beta}^{i_k} \equiv \sum_{j_k} \tilde{U}_{\alpha,\beta}^{i_k,j_k} \langle j_k| \psi_I^{[k]}\rangle$
for the resulting isometry $\tilde{V}_{[k]}^{\A\B}$.
In the ideal case, when the fidelity reaches unity, the ancilla can be set to decouple unitarily in the last step. However, this will not be the case in general when the allowed ancilla-qubit unitaries are restricted. Thus, the optimization problem reads
\begin{eqnarray}
\label{eq:cost_func}
\min_{|\tilde{\Psi}\rangle {\in \tilde{H}_{[k]}} }
\parallel |\tilde{\Psi}\rangle - |\varphi_F\rangle \otimes |\psi\rangle  \parallel^{2}  \;,
\end{eqnarray}
involving a multivariable cost function in $|\varphi_F\rangle$ and $\{\bar{h}_1^{[n]},\dots,\bar{h}_1^{[1]}\}$,
with $\bar{h}_1^{[k]}=h_1^{[k]} t$, as the \emph{variational parameters}, which can be solved in an iterative procedure.
We start by picking a particular
unitary, say $\tilde{U}_{[k]}^{\A\B}$, and minimizing the cost function in~\Eq{eq:cost_func}, varying over $\bar{h}_1^{[k]}$, and regarding couplings of all the other
unitaries as fixed. Then we move on to the neighboring unitary and optimize its coupling. When
all unitaries have been optimized locally, we sweep back again and so forth until convergence.
Each iteration of the local optimization procedure requires the calculation of the overlap of the
states in the cost function of~\Eq{eq:cost_func}, which can be straightforwardly
calculated in MPS representation as illustrated in Fig.~\ref{opt_pat} (with $\UUU^{\A}$ and $\UUU^{\B}$ set to $\mathds{1}$ there). Varying over the vector $|\varphi_F\rangle$ and using the resulting optimal one, the cost function simplifies to $2(1-\| \langle \tilde{\Psi} | \psi\rangle \|)$, suggesting the definition of the fidelity of the procedure as $\F \equiv \| \langle \tilde{\Psi} | \psi\rangle \|$.

\begin{figure}[t]
\centering
\includegraphics[width=1\linewidth]{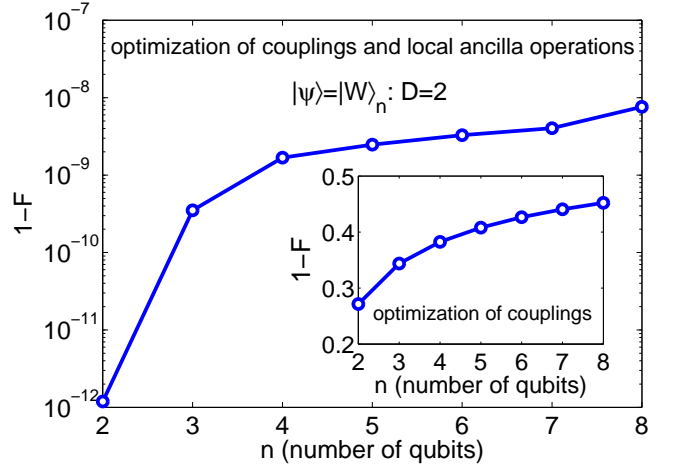}
\caption{(Color online) The deviation of the fidelity $1 - \F=1 - \| \langle \tilde{\Psi} | \psi\rangle \|$ as a function of the number $n$ of qubits for the $W$ state with $D=2$
when optimizing the couplings $h_{j_{\A}j_{\B}}$ and the local ancilla unitaries $\UUU_{A}$, with initial qubit states
all equal, $|\psi^{[k]}_I\rangle=|0\rangle$. The inset shows the case where only the couplings $h_{j_{\A}j_{\B}}$ are being optimized.
}
\label{opt}
\end{figure}

For the restricted entangling Hamiltonian of \Eq{eq:restricted_Hamiltonian}, the variational space is so small (only one parameter at each step), that the variational optimization
procedure in general does not result in much overlap with the target state $|\psi\rangle$, as illustrated
in the inset of Fig.~\ref{opt} using the familiar $|W\rangle_n$ state as target.
However, $\F$ can be improved by enlarging the variational space. For example, consider $\tilde{U}^{\A\B}$ in \Eq{eq:joint_psi} being
replaced with restricted unitaries of the form $\UUU_{[k]}^{\A} \tilde{U}_{[k]}^{\A\B}$,
where $\UUU_{[k]}^{\A}=e^{-iH_{[k]}^{\A} t/\hbar}$ are arbitrary local ancilla unitaries of dimension $D \times D$.
This optimization problem can be treated in the same manner as the one described in~\Eq{eq:cost_func},
except that before optimizing each $\tilde{U}^{\A\B}$, we will also vary over the ancilla operation~$\UUU^{\A}$. In this way, we are able
to produce the $|W\rangle_n$-state with almost perfect fidelity (e.g., $1-\F \approx 10^{-9}$ for $n=4$)
as illustrated in Fig.~\ref{opt}. In both cases, the smaller the number of qubits $n$, the larger the fidelity,
which is a purely numerical issue due to the local optimization.
Models requiring the entangling Hamiltonian of the $XXZ$ form $h_1^{[k]} (\sigma_1 \otimes \sigma_1+\sigma_2 \otimes \sigma_2)+h_2^{[k]} \sigma_3 \otimes \sigma_3$, can be simulated in a similar manner. %\tilde{H}_{[k]}^{\A\B}=

%Moreover, we have found strong numerical evidence that an arbitrary MPS with $D=2$ can be generated by considering
%$\tilde{U}^{\A\B}$ in \Eq{eq:joint_psi} (parametrized by $\bar{h}_1^{[k]}$) replaced by restricted unitaries of the
%form $\UUU^{\A}_{[k]} \UUU^{\B_I}_{[k]} \tilde{U}_{[k]}^{\A\B} \UUU^{\B_F}_{[k]}$ where $\UUU_{[k]}^{\B}=e^{-iH_{[k]}^{\B} t/\hbar}$ are %arbitrary local qubit unitaries (see Fig.~\ref{opt_pat}).

\begin{table*}[t] 
\caption{Comparing the optimal couplings of our simulation $h_1^{\text{sim}}$ to those used for experimental realization of $W$ state
$h_1^{\text{expt}}$ in~Ref.~\cite{Haf05} for $n=5$.}
\centering
\begin{ruledtabular}
\begin{tabular}{c c c c c c}
%\hline\hline
Site index ($k$) &1 &2 &3 &4 &5 \\ %[.5ex]
\hline \\ \vspace{2mm}
$[({h_1^{\text{sim}}/h_1^{\text{expt}}}) -1] \times 10^5$ & 36.50 & 0.72 & 8.64 & 0.62 & 0.59\\  [.5ex]
%heading
%\hline
%1 & $1.221 \times 10^{-5}$ \\
%2 & $1.541 \times 10^{-6}$ \\
%3 & $2.521 \times 10^{-5}$ \\
%4 & $1.252 \times 10^{-6}$ \\
%5 & $1.850 \times 10^{-6}$ \\
%6 & $8.751 \times 10^{-6}$ \\ [1ex]
%\hline
\end{tabular}
\end{ruledtabular}
\label{table:Blatt}
\end{table*}

As a test of the proposed protocols, we applied our variational prescription to the sequential generation of $W$ states in an ion chain. Following closely the recent experiment of~Ref.~\cite{Haf05}, we targeted a $W$ state with the entangling Hamiltonian of the form
$h_1 (\sigma^{+} \otimes \sigma^{+}+\sigma^{-} \otimes \sigma^{-})$, with $\sigma^+$ and $\sigma^{-}$ being the usual raising and lowering Pauli operators, respectively, and the initial state $|\psi_I\rangle=|1\rangle |0\rangle \dots |0\rangle$ used in experiment.
The optimal couplings $h_1^{\rm{sim}}$ of the resulting converged variational MPS $|\tilde{\Psi}\rangle$ (with $1-\F \approx 10^{-9}$ for $n=5$) turned out to agree very well with the two-qubit rotations $h_1^{\rm{expt}}$ used for the experiment of Ref.~\cite{Haf05}, as illustrated in Table~\ref{table:Blatt}.

As the main result of this paper, we have found strong numerical evidence that an \emph{arbitrary} MPS with $D=2$ can be generated sequentially if the single-parameter restricted unitaries $\tilde{U}^{\A\B}$ in \Eq{eq:joint_psi} [based on \Eq{eq:restricted_Hamiltonian}] are augmented by arbitrary local unitaries for both ancilla and qubit spaces. The combined unitary employed was $\UUU^{\A}_{[k]} \UUU^{\B_I}_{[k]} \tilde{U}_{[k]}^{\A\B} \UUU^{\B_F}_{[k]}$,
where $\UUU_{[k]}^{\B}=e^{-iH_{[k]}^{\B} t/\hbar}$ are arbitrary local qubit unitaries (see Fig.~\ref{opt_pat}).
We have considered, for this purpose, the generation of 100 randomly chosen MPS and have found that $1-\F$ remains below $6 \times10^{-13}$ up to $n=5$. Note that the combined action of these unitaries includes (at most) 11 real independent parameters, which in practice can be reduced to
ten, since varying a global phase has no effect. In contrast, the unrestricted unitaries $U^{\A\B}$ involve 16 real independent
parameters. Thus sequential generation of an arbitrary MPS with $D=2$, can be achieved more economically than previously
realized: a sufficient condition is the availability of the set of restricted two-qubit \emph{isometries} specified above, instead of the availability of arbitrary two-qubit \emph{unitaries}~\cite{Schoen_PRL}.

We may then wonder whether some fixed parameter-free two-qubit isometries can act as universal set for generation of arbitrary entangled states.
The problem we propose, which is the natural one in the sequential
generation of multi-qubit states, is the following: give a minimal set $S$
of two-qubit unitaries such that one can generate an arbitrary isometry
with a single unitary of the set $S$, together with arbitrary one-qubit
unitaries. Note that, we already showed numerically that $S$ can be given by the
single-parameter interactions of the $XY$ type, whereas we now wonder whether this
can be realized by a minimal set of fixed canonical gates. Note that since the paradigm is completely different
(a single use of the entangling unitary and isometries instead of
unitaries), the results concerning universal sets of gates for quantum
computing do not play a role for our protocol.
We have found numerically, for example, that some parameter-free fixed two-qubit gates [such as controlled NOT (CNOT)] plus three local unitaries are not {\it isometrically} universal, as they are not capable of generating an arbitrary state with $\F=1$.
The search for such two-qubit gates, if any, remains open.

Recently, a lot of effort has been devoted to find minimal sets of one-qubit and two-qubit gates, and the minimal number of applications, to generate arbitrary two-qubit {\it unitaries}~\cite{Whaley05}. The existence of these universal sets is of central relevance in quantum computing. The above results suggest consideration of a class of problems involving a different paradigm: which are the universal sets of one-qubit and two-qubit gates that can generate arbitrary two-qubit {\it isometries}? What is the minimal number of applications and how does this compare to the quantum computing case?
For the case of two-qubit unitaries, a universal gate set (in the usual quantum computing sense) is clearly
sufficient, but not necessary. This results, for example, from counting
the number of independent parameters for an arbitrary two-qubit unitary,
clearly larger than in the case of an arbitrary two-qubit isometry.
The aim will be then to find the exact decomposition of an arbitrary isometry into a minimal applications of unitaries as computational primitives. The general solution associated with this paradigm remains open.

Finally, we also want to point out that our scheme by
construction can be clearly viewed also within the
general framework of optimal control theory~\cite{Glaser05,Timoney}.

\section{Conclusions}
\label{sec:conclusions}

In conclusion, we have developed protocols for an efficient sequential generation of entangled multiqubit states under realistic experimental
constraints. We stress that the proposed optimization methods are of wide applicability and will be of importance for any sequential physical setup. In particular, we can mention photonic qubits, atoms, ions, superconducting qubits, or quantum dots.

%We point out the current problem may also be studied with optimal control theory schemes developed in Reference~\cite{Glaser05}.
%in which efficient algorithms for quantum simulations in networks of coupled qubits are offered.

%It is noteworthy to mention that the realization of single qubit gates with trapped ions that are robust against experimental imperfections has been %investigated in Ref.~\cite{Timoney} by using optimal control theory techniques.

\begin{acknowledgments}

H. S. and D.P.-G. thank Universidad del Pa\'{\i}s Vasco for hospitality.
H.S., A.W., and J.v.D. acknowledge support from Spintronics RTN, the DFG
(SFB 631, De-730/3-2), and the German
Excellence Initiative via the Nanosystems Initiative Munich (NIM). L.L. thanks funding from Alexander von Humboldt Foundation and support from MEC Project No. FIS2008-05705/FIS. D. P.-G. acknowledges support from Spanish projects MTM2005-00082 and CCG07-UCM/ESP-2797, and E.S. from Ikerbasque Foundation, EU EuroSQIP project, and UPV-EHU Grant No. GIU07/40.

\end{acknowledgments}


\begin{thebibliography}{10}


\bibitem{Nielsen_Chuang}
M.~A. Nielsen and I.~L. Chuang,
\emph{Quantum Computation and Quantum Information}
(Cambridge University Press, Cambridge, England, 2000).

\bibitem{Schoen_PRL}
C. Sch\"on, E. Solano, F. Verstraete, J.~I. Cirac, and
M.~M. Wolf,  Phys. Rev. Lett. \textbf{95}, 110503 (2005).

\bibitem{Schoen_PRA}
C. Sch\"on, K. Hammerer, M.~M. Wolf, J.~I. Cirac, and E. Solano,
Phys. Rev. A \textbf{75}, 032311 (2007).

\bibitem{David}
D. P\'erez-Garc\'ia, F. Verstraete, M.~M. Wolf, and J.~I. Cirac,
Quantum Inf. Comput. \textbf{7}, 401 (2007).

\bibitem{AKLT+Fannes} M. Fannes, B. Nachtergaele, and R.~F. Werner,
Commun. Math. Phys. \textbf{144}, 443 (1992).

\bibitem{DMRG} S. R. White, Phys. Rev. Lett. \textbf{69}, 2863 (1992);
U. Schollw\"ock, Rev. Mod. Phys. \textbf{77}, 259 (2005).

\bibitem{DMRG_MPS} S. \"Ostlund and S. Rommer, Phys. Rev. Lett.
\textbf{75}, 3537 (1995).
%; J. Dukelsky, M.~A. Mart\'in-Delgado, T. Nishino, and G. Sierra,
%Europhys. Lett. \textbf{43}, 457 (1998).

%\bibitem{Dukelsky}
%J. Dukelsky, M.~A. Mart\'in-Delgado, T. Nishino, and G. Sierra,
%Europhys. Lett. \textbf{43}, 457 (1998).

\bibitem{Krishna} H.~R. Krishna-Murthy, J.~W. Wilkins, and K.~G. Wilson,
Phys. Rev. B \textbf{21}, 1003 (1980).

\bibitem{SWV_cloning} H. Saberi, A. Weichselbaum, and J. von Delft,
Phys. Rev. B \textbf{78}, 035124 (2008).

\bibitem{GHZ}
D.~M. Greenberger, M. Horne, and A. Zeilinger, in
\emph{Bell's Theorem, Quantum Theory, and Conceptions of the
Universe}, edited by M. Kafatos (Kluwer, Dordrecht, 1989).

\bibitem{Warticle}
W. D\"ur, G. Vidal, and J. I. Cirac, Phys. Rev. A {\bf 62}, 062314 (2000).

\bibitem{Cluster}
R. Raussendorf and H. J. Briegel, Phys. Rev. Lett. {\bf 86}, 5188 (2001).

%\bibitem{Pan00} J.-W. Pan, D. Bouwmeester, M. Daniell, H. Weinfurter, and A. Zeilinger, Nature {\bf 403}, 515 (2000).
\bibitem{Pan00} J.-W. Pan {\it et al.}, Nature {\bf 403}, 515 (2000).

\bibitem{Lei05} D. Leibfried {\it et al.}, Nature {\bf 438}, 639 (2005).

\bibitem{Haf05} H. H\"affner {\it et al.}, Nature {\bf 438}, 643 (2005).

\bibitem{Rau00} A. Rauschenbeutel {\it et al.}, Science {\bf 288}, 2024 (2000).

\bibitem{Kie07} N. Kiesel, C. Schmid, G. T\'oth, E. Solano, and H. Weinfurter, Phys. Rev. Lett. {\bf 98}, 063604 (2007).

\bibitem{Kie05} N. Kiesel, C. Schmid, U. Weber, G. Toth, O. Guhne, R. Ursin, and H. Weinfurter, Phys. Rev. Lett. {\bf 95}, 210502 (2005).

\bibitem{ConstrainedOptimizationBook} D. P. Bertsekas, {\it Constrained Optimization and Lagrange Multiplier Methods} (Athena Scientific, Belmont, MA, 1996).

\bibitem{SVD_image}
H.~C. Andrews and C.~L. Patterson, IEEE Trans. Commun. \textbf{24},
425 (1976).

\bibitem{Golub_VanLoan}
G.~H. Golub and C.~F. Van Loan,
\emph{Matrix Computations}
(The Johns Hopkins University Press, Baltimore, 1996).

\bibitem{Horn_Johnson}
R.~A. Horn and C.~R. Johnson,
\emph{Topics in Matrix Analysis}
(Cambridge University Press, Cambridge, England, 1991).

\bibitem{VPC} F. Verstraete, D. Porras, and J.~I. Cirac, Phys. Rev.
Lett. \textbf{93}, 227205 (2004).


\bibitem{Vidal}
G. Vidal, Phys. Rev. Lett. \textbf{91}, 147902 (2003).

\bibitem{Whaley03}
K.~R. Brown, J. Vala, and K.~B. Whaley, Phys. Rev. A \textbf{67}, 012309 (2003).


\bibitem{Whaley05}
See, for example, J. Zhang, and K.~B. Whaley, Phys. Rev. A \textbf{71}, 052317 (2005), and references therein.

\bibitem{Glaser05}
T. Schulte-Herbr\"{u}ggen, A. Sp\"{o}rl, N. Khaneja, and S. J. Glaser, Phys. Rev. A \textbf{72}, 042331 (2005).

\bibitem{Timoney}
N. Timoney, V. Elman, W. Neuhauser, and C. Wunderlich, Phys. Rev. A \textbf{77}, 052334 (2008).


\end{thebibliography}
\end{document}